\documentclass[sn-aps]{sn-jnl}
\usepackage{amsmath}
\usepackage{accents}
\usepackage{graphicx}
\usepackage{bm}
\usepackage{epstopdf}
\usepackage{float}
\usepackage{dcolumn}
\usepackage{bm}
\usepackage{mathrsfs}
\usepackage{amsmath}
\usepackage{footnote}
\usepackage{accents}
\usepackage{tensor}
\usepackage{mathtools}
\usepackage[utf8]{inputenc}
\usepackage[T1]{fontenc}
\usepackage{lmodern}

\jyear{2024}%

\theoremstyle{thmstyleone}%

\theoremstyle{thmstyletwo}%

\theoremstyle{thmstylethree}%

\begin{document}

\title[The Role of Torsion in Trans-Planckian Physics]{The Role of Torsion in Trans-Planckian Physics}

\author[1]{\fnm{Elham} \sur{Arabahmadi}}\email{Elhaam.arabahmadi@gmail.com}

\author*[1]{\fnm{Siamak} \sur{Akhshabi}}\email{s.akhshabi@gu.ac.ir}

\affil[1]{\orgdiv{Department of Physics, Faculty of Sciences}, \orgname{ Golestan University}, \orgaddress{ \city{Gorgan}, \country{IRAN}}}

\abstract{The torsion of spacetime, if it exists, plays an important role at the very early universe, when the spin density of particles was very high. It is generally believed that in the extremely high energies of the early universe a new physics called trans-Planckian physics should be considered. Since the initial conditions for inflation are probably a result of this new physics, here we consider spin and torsion as trans-Planckian effects and analyze their influence on the power spectrum of scalar and tensor perturbations at the end of the inflationary era.}


\maketitle

\section{Introduction}\label{sec1}

Despite the success of general relativity and the standard model of cosmology it appears that the effects of a new physics, generally accepted to be the elusive theory of quantum gravity, should be taken into consideration at the very high energies of the early universe, near to the big bang. These effects are conveniently called trans-Planckian as they become very important at energy scales close or higher than the Planck scales. Although there is still no clear agreement on the exact nature of this new trans-Planckian physics, it is expected that it can find an answer to the important problem of compatibility of gravity with quantum mechanics.
The effects of this trans-Planckian physics can change the power spectrum of initial density fluctuations produced at the end of the inflationary period, because the initial conditions at the beginning of that period are probably affected by the trans-planckian effects \cite{Niemeyer, Hui, Shiu, Shankar, Kempf, Easther2, Kaloper, Brandenberger1}. On the other hand, recent analysis of the current inflationary models seems to suggest that the initial energy scale of inflation should be significantly lower than the Planck scale  \cite{BrandennVafa,YongCai}.
More specifically, it has previously been realized that if the inflationary phase lasts long enough to successfully produce the cosmological structures observed today, the present-day length scales should originate from modes that are smaller than the Planck length during inflation \cite{Martin, BrandennMartin, Niemeyer1}. This is the so-called trans-Planckian problem of the inflationary cosmology \cite{Chalmers, Easther1, Easther3, Bozza, MartinnBranden}.
It was generally expected that a complete theory of quantum gravity would avoid this problem; however, it has been argued in \cite{BedroyanVafa}  that in a consistent quantum theory of gravity, sub-Planckian quantum fluctuations should remain quantum and never become larger than the Hubble horizon. This is called the Trans-Planckian Censorship Conjecture (TCC), which forces the trans-Planckian perturbation modes to be hidden by the Hubble horizon.
The combination of the trans-Planckian problem and the TCC puts strict bounds on some inflationary parameters in models with single or multiple scalar fields, most specifically the number of e-folds, energy density of the inflation, slow-roll parameters and the tensor-to-scalar ratio \cite{BrandennVafa}. There has been considerable interest in the literature on the implications of these issues and their resolution; see, for example, \cite{Broy, Shojai1, Kadota, Brahma, Tenkanen, YongCai2, Torabian, KamalinBranden, WilsonnBranden, Sasaki1}. 

Most of the analysis mentioned above is based on the assumption that the theory of general relativity (GR) and the standard model of cosmology, which is based on GR, can correctly describe the evolution of the universe, at least from a few seconds after the big bang to the present day. However  various issues in the standard model of cosmology, such as the singularity problem and the nature of dark matter and dark energy, led researchers to consider gravity theories that modify general relativity in some way.
These modified gravity theories have attracted considerable interest in recent times and various aspects of them have been studied extensively in the literature \cite{Faraoni, Nojiri1, Cappozz1, Clifton, Nojiri2, Odintsov}. The issue of the trans-Planckian problem in modified gravities is also of great interest, as these theories could potentially modify the predictions and constraints of the inflationary era. As an example, trans-Planckian corrections to the power spectrum of scalar perturbations in $f(R)$ gravity have been recently studied in \cite{Shojai2}.
Among the modified theories of gravity, the ones containing torsion are some of the most natural extensions of GR \cite{Blag1, Hayashi}. Torsion, which is coupled to the spin of matter, appears naturally in many theories of quantum gravity \cite{Scherk, Hehl1974, Hehl1976, Shapiro}. In particular, it has been shown that at low energies, the effective Lagrangian of string theory is mathematically equivalent to a metric theory of gravity that includes torsion \cite{Hammond1, Hammond2}. Additionally, the strength of the field associated with the Kalb-Ramond field in string theory behaves as a torsion field within the context of field geometry \cite{Kalb}.
 Penrose has also demonstrated that torsion naturally emerges when spinors are permitted to undergo rescaling through a complex conformal factor \cite{Penrose}.

Therefore, it seems that torsion effects, if they exists, are important in the Trans-planckian era and their effects can potentially be seen in the observations related to the time close to the big bang, especially the inflation period. In this article, considering the space-time torsion as a trans-planckian effect, we investigate its influence on cosmic parameters, especially inflationary parameters. To estimate the Trans-Planckian effects of torsion, we use the Einstein-Cartan theory (ECT), which is the simplest generalization of general relativity in the presence of torsion.
In 1922, Cartan introduced his theory to suggest that torsion represents the macroscopic expression of the intrinsic angular momentum (spin) of matter \cite{Cartan}. Several years later, the concept of  spin was reintroduced independently into general relativity by Kibble and Sciama.\cite{Kibble}.
Subsequently, the EC theory has been studied extensively as it became the subject of much research; (for example see \cite{Hehl1976, Trautman} and references therein) as it serves as the most straightforward classical extension of Einstein’s general relativity \cite{Blagojevic, Battista1, Battista2}. Generalizations of Einstein-Cartan theory also include the Poincar{\'e} gauge theory of gravity \cite{Obukhov1} and the most general metric-affine theory of gravity \cite{Hehl1995}. Indeed, there are several avenues for experimentally testing gravitational theories that incorporate non-vanishing spacetime torsion (see for example \cite{Hammond,Mao,Kostelecky,March,Hehl2013,Puetzfeld,Lin, Zamani1, Pereira, Akhshabi, Ranjbar}). 
However, experimental or observational evidence validating the unique predictions of theories with torsion or confirming the existence of spacetime torsion remains elusive. The primary reason for this absence of evidence lies in the fact that these theories depart from classical general relativity only at exceptionally high energy densities \cite{Kranas}. Such extreme densities can only be found in extremely compact objects like neutron stars and black holes, or during the initial stages of the universe’s evolution, the latter of which we focus on this article.

There are various ways to approach the trans-Planckian problem in inflationary cosmology. One way, which we follow here is based on the choice of the vacuum state and the time to impose the initial condition for inflation. If we had been allowed to follow a given mode arbitrarily far back in time, we could have argued for a unique vacuum, the Bunch-Davies vacuum, which also is the vacuum typically used in calculations of the primordial spectrum. In the presence of a fundamental scale,  one cannot follow a given fluctuation mode to scales smaller than the Planck length \cite{Danielsson1, Danielsson2, Danielsson3}.
As a result, in order to incorporate the effects of trans-Planckian physics, which now include significantly large torsion and spin close to the initial singularity, we assume that all effects due to unknown high-energy quantum gravity theory can be captured by a suitable choice of vacuum and then look for signs of the new trans-Planckian physics in fluctuations of the cosmic microwave background or in large-scale structures \cite{Danielsson4, Cielo}. This assumption is particularly natural in an expanding universe, as has been elaborated in \cite{Danielsson2}. The choice of having a non-Bunch-Davies vacuum has also been studied in the context of the swampland in string theory \cite{Ashoorioon, Wali}. 

In subsequent sections we use this procedure to calculate various  parameters at the end of the inflationary era. structure of the paper is as follows: In section \ref{sec2} we review the fundamental equations of Einstein-Cartan theory of gravity and present the Friedmann equations used to extract the cosmological dynamics in this theory. Section \ref{sec3} is devoted to studying the scalar perturbations generated at the end of inflation, taking into account the effects of spin and torsion on their spectrum.
In section \ref{sec4}, by using the results of previous section, we examine the effects of the new choice of vacuum in trans-Planckian physics to determine the modified power spectrum in the presence of torsion. Section \ref{sec5} briefly discusses the generation and propagation of tensor perturbations (i.e. gravitational waves) in Einstein-Cartan cosmology. Finally, section \ref{sec6} is devoted to the discussion and conclusions. 

\bigskip
\section{Einstein-Cartan Theory of Gravity}\label{sec2}

In this section we mainly follow the conventions and notation of reference \cite{mainEC}. The Lagrangian in the Einstein-Cartan theory of gravity is assumed to be the same as the Einstein-Hilbert action of general relativity.

\begin{equation}
S=\frac{1}{2\kappa} \int d^{4}x \mathfrak{R}+\int d^{4}x \mathcal{L}_{m}
\end{equation}

Here $\mathfrak{R}$ is the Ricci scalar constructed with the full connection, which now contains torsion and curvature. As a result, there are two independent dynamical variables: the metric tensor as well as the new independent torsion field (which can also be described by the independent part of the spin connection), both of these fields contributes to the total gravitational interactions. In the subsequent calculations, we assume the metric signature to be (+,-,-,-). Torsion tensor is defined as the antisymmetric part of the connection 

\begin{equation}
\label{1}
T_{\mu \nu}^\alpha =\Gamma_{[\mu \nu]}^\alpha = \frac{1}{2}(\Gamma_{\mu \nu}^\alpha- \Gamma_{\nu \mu}^\alpha)
\end{equation}

We also defined the modified torsion tensor as

\begin{equation}
Q\indices{_{\mu\nu}^{\alpha}}\equiv T\indices{_{\mu\nu}^{\alpha}}+\delta_{\mu}^{\alpha}T\indices{_{\nu\lambda}^{\lambda}}-\delta_{\nu}^{\alpha}T\indices{_{\mu\lambda}^{\lambda}}
\end{equation}

In ECT, the connection can be divided into two parts

\begin{equation}
\label{2}
\Gamma_{\mu \nu}^\alpha = \overset {\circ}{\Gamma}^\alpha _{\mu \nu} - K_{\mu \nu}^\alpha
\end{equation}

where $\overset {\circ}{\Gamma}\,^\alpha _{\mu \nu}$ is the Christoffel connection and $ K_{\mu \nu}\,^\alpha$ is called the contorsion tensor. The relation between contorsion and torsion tensors can be expressed as

\begin{equation}
\label{3}
K_{\mu \nu}\, ^\alpha = -T_{\mu \nu}\, ^\alpha + T_\nu \,^ \alpha \, _ \mu - T^\alpha \, _{\mu \nu}
 \end{equation}
 
\begin{equation}
\label{4}
T_{\mu \nu}\, ^\alpha = - K_{[\mu \nu]}\, ^\alpha = -\frac{1}{2}\,(K_{\mu \nu}\, ^\alpha - K_{\nu \mu}\, ^\alpha)
\end{equation}

Field equations of Einstein-Cartan gravity are obtained by varying the action with respect to dynamical variables \cite{Hehl1976,Hehl1974,Hehl1973,Hehl19744}. For the matter part, the metric energy momentum tensor $\sigma_{\mu\nu}$ and the spin tensor $S_{\,\,\nu\mu}^\alpha$ are defined as \cite{mainEC}

\begin{equation}
\label{EMTensors}
\sigma_{\mu\nu}:=-\frac{2}{\sqrt{-g}}\frac{\delta \mathcal{L}_{m}}{\delta g^{\mu\nu}}\mbox{ }\mbox{ } \mbox{ , } \mbox{ } \mbox{ } S\indices{_{\alpha}^{\nu\mu}}:=\frac{1}{\sqrt{-g}}\frac{\delta \mathcal{L}_{m} }{\delta K\indices{_{\mu\nu}^{\alpha}}}
\end{equation}

Using the above relations, the two field equations of Einstein-Cartan theory will have the following form \cite{mainEC}

\begin{equation}
\label{EinsteinHehl0}
G_{\mu\nu}+\accentset{\star}{\nabla}_{\lambda}\left(-Q\indices{_{\mu\nu}^{\lambda}}+Q\indices{_{\nu}^{\lambda}_{\mu}}-Q\indices{^{\lambda}_{\mu\nu}} \right)=\kappa \sigma_{\mu\nu}
\end{equation}

\begin{equation}
\label{TorsionHehl}
T\indices{_{\lambda\nu}^{\mu}}+\delta_{\lambda}^{\mu}T\indices{_{\nu\alpha}^{\alpha}}-\delta^{\mu}_{\nu}T\indices{_{\lambda\alpha}^{\alpha}}=\kappa S\indices{_{\lambda\nu}^{\mu}}
\end{equation}

where we defined $\accentset{\star}{\nabla}_{\lambda}=\nabla_{\lambda}+2T_{\lambda \alpha}^{~~\alpha}$. The second field equation \eqref{TorsionHehl} is just an algebraic equation for torsion in terms of its source $S\indices{_{\mu\nu}^{\alpha}}$ while by some manipulations, the first equation, i.e. \eqref{EinsteinHehl0} can be brought to the following form resembling the usual Einstein equation \cite{mainEC}

\begin{equation}
\label{EinsteinHehl}
\accentset{\circ}{G}_{\mu\nu}\equiv \accentset{\circ}{R}_{\mu\nu}-\frac{1}{2}g_{\mu\nu}\accentset{\circ}{R}=\kappa \tilde{\sigma}_{\mu\nu}
\end{equation}

Here, $\tilde{\sigma}_{\mu\nu}$ will include the torsion source terms 

\begin{equation}
\label{HehlSigma}
\begin{split}
\tilde{\sigma}_{\mu\nu}:=\sigma_{\mu\nu}+\kappa & \left\{-4S\indices{_{\mu\lambda}^{[\alpha}}S\indices{_{\nu\alpha}^{\lambda]}}-2S_{\mu\lambda\alpha}S\indices{_{\nu}^{\lambda\alpha}}+S_{\alpha\lambda\mu}S\indices{^{\alpha\lambda}_{\nu}} \right. \\ 
& \left. +\frac{1}{2}g_{\mu\nu}\left(4S\indices{_{\lambda}^{\beta}_{[\alpha}}S\indices{^{\lambda\alpha}_{\beta]}}+S^{\alpha\lambda\beta}S_{\alpha\lambda\beta} \right) \right\}
\end{split}
\end{equation}

In order to examine the cosmological dynamics in Einstein-Cartan theory, we need to examine spatially homogeneous and isotropic spacetimes with non-vanishing torsion. In the cosmological context,  one may associate torsion with the intrinsic spin of the matter components in the universe, or it may also be treated as a geometric entity intrinsic to the background spacetime \cite{Capozziello,Kranas}. In the latter case, the symmetries of the Friedmann-Robertson-Walker spacetime force the torsion tensor to have a very specific form which preserves the spatial homogeneity and the spatial isotropy of the background \cite{Tsamparlis, Olmo, Beltran}. In the most general case, the torsion is fully determined by two scalar functions depending only on time \cite{Goenner}. However, in the  Einstein-Cartan theory where Weyssenhoff fluid is usually used to describe spin fluid \cite{Weyssenhoff, Obukhov}, a single time dependent function is sufficient to completely describe the torsion tensor. The corresponding  non-zero components of the spin tensor then can be found from the algebraic field equation (\ref{TorsionHehl}) as

\begin{equation}
\label{spint}
s(t)=S_{123}=S_{231}=S_{312}=-S_{[123]}=S_{0ii}=S_{022}=S_{033}=-S_{i0i}
\end{equation}

By making these choices, we can now derive the torsional counterparts of the familiar Friedmann equations by substituting the Friedmann-Robertson-Walker metric and the spin tensor (\ref{spint}) into the field equations (\ref{EinsteinHehl}) and assuming that the usual energy-momentum tensor for matter in the universe $\sigma_{\mu\nu}$ is described by a perfect fluid. The final result is given in reference \cite{mainEC} as

\begin{equation}
\label{ECfriedmann}
H^2=\frac{\kappa }{3}\,(\rho-\frac{1}{4}\kappa s^2)
\end{equation}

\begin{equation}
\label{ECAcc}
H^{2}+\dot{H}=-\frac{\kappa }{6}(\rho+3\,p-\kappa s^2)
\end{equation}

In which $\kappa$ is the coupling constant, $s$ is the spin density and $\rho$ and $p$ are the energy density and isotropic pressure of the fluid, respectively. As can be seen from the form of the Friedmann equation \ref{{ECfriedmann}}, the term involving the energy density $\rho$ is proportional to the coupling constant $\kappa$, however, the spin term is proportional to  $\kappa^{2}$ and as a result the effects of spin and torsion on the evolution of the Universe
is significant only at very large densities of matter available at the very early universe. Assuming that the spins of fermionic matter in the universe are not polarized, the spin density $s$ can be related to the energy density of matter $\rho$ and subsequently to the scale factor $a(t)$ by relation \cite{Gasp, Nurg, mainEC, Qorani}

\begin{equation}
\label{spinsquar}
\begin{split}
s^2 &=\frac{1}{2} \langle S_{ijk} S^{ijk} \rangle =\frac{1}{8} (\hbar c)^2 \frac{n_0^2}{\rho_0^{2/(1+\omega)}} \rho^{2/(1+\omega)}\\
& = B_\omega \rho^{2/(1+\omega)} = B_\omega \rho_0^{2/(1+\omega)} a^{-6} 
\end{split}
\end{equation}

that $S_{ijk}$ is the spin tensor, $\omega$ is the equation of state for the matter in the universe, $\rho_0$  and $n_0$  are the present day energy density and particle number density respectively,  and $B_{\omega}$ is a dimensional constant depending on $\omega$. In the next section we will use the Friedmann equations and the expression for the spin density mentioned above to examine the effects of spin and torsion on the spectrum of scalar perturbations at the end of inflation.

\bigskip
\section{The Effects of Torsion On the Spectrum of Scalar Perturbations  }\label{sec3}

To begin with, we consider a perturbed scalar field (inflaton) as follows

\begin{equation}
\label{inflaton}
\phi(\tau,x)\,=\,\bar{\phi}(\tau)\,+\,\delta\phi(\tau,x)\,=\,\bar{\phi}(\tau)\,+\,\frac{f(\tau,x)}{a(\tau)}
\end{equation}

where $\phi(\tau,x)$ represents the perturbed scalar field, $\bar{\phi}(\tau)$ is the unperturbed background field and $f(\tau,x)$ is the perturbation function. In order to see how perturbations changes in presence of torsion, we shall start with the action of the inflaton field 

\begin{equation}
\label{action}
S=\int d\tau  d^3 x\sqrt{-g}\,\bigg[\frac{1}{2}g^{\mu\nu}\partial_\mu\phi\partial_\nu\phi-V(\phi)\bigg]\,.
\end{equation}

where $g \equiv det(g_{\mu \nu})=-a^8$ and $\tau$ is the conformal time. We consider the unperturbed FRW line element as $ds^2=a^2(\tau)(d\tau^2-dr^2)$. Using this metric and relation (\ref{inflaton}) in action (\ref{action}), we will have

\begin{equation}
\begin{split}
S&=\int d\tau  d^3x \sqrt{-g}\,\bigg[\frac{1}{2}g^{\mu\nu}\partial_{\mu}\phi\partial_{\nu}\phi-V(\phi)\bigg]\\
&=\int d\tau d^3x \bigg[\frac{a^2}{2}\Big((\phi^\prime)^2-(\nabla\phi)^2\Big)-a^4V(\phi)\bigg]
\end{split}
\end{equation}

where a 'prime' denotes differentiation with respect to the conformal time\footnote{In general in this article a 'prime' denotes differentiation with respect to the argument of the function}. For the perturbed scalar field (\ref{inflaton}), the action will be

\begin{equation}
\label{Perturbed action}
\begin{split}
S & =\int d\tau d^3x \bigg [\frac{1}{2}a^2 \bigg (\Big (\frac{\partial(\bar{\phi}(\tau)\,+\,\frac{f(\tau,x)}{a(\tau)})}{ \partial\tau}\Big)^2 \\
& -\, \Big(\nabla\big(\bar{\phi}(\tau)\,+\,\frac{f(\tau,x)}{a(\tau)}\big)\Big)^2\bigg)\,-\, a(\tau)^4\,V(\phi)\bigg]\\
\end{split}
\end{equation}

To extract the linearized perturbation dynamics, we will need the quadratic action \cite{Baumann}. By performing the Taylor expansion for the potential $V(\phi))$ in (\ref{Perturbed action}) and focusing on the terms which are second order in perturbation function $f$, we get the quadratic action after some straightforward algebra as

\begin{equation}
\label{quadAc}
\begin{split}
S^{(2)} &=\int d\tau d^3x\bigg[\frac{1}{2}a^2(\tau)\bigg(\big(\frac{f^\prime (\tau,x)}{a(\tau)}\big)^2-2\frac{f^\prime(\tau,x) a^\prime(\tau)f(\tau,x)}{a^3(\tau)}\\
& +\big(\frac{a^\prime(\tau) f(\tau,x)}{a^2(\tau)}\big)^2-\big(\frac{\nabla f(\tau,x)}{a(\tau)}\big)^2\bigg)-a^4(\tau)\frac{V^{\prime\prime} (\phi)}{2!}\frac{f^2(\tau,x)}{a^2(\tau)}\bigg]\\
& =\int d\tau d^3x \bigg(\frac{1}{2}f^{\prime2}(\tau,x) - \frac{a^\prime(\tau)}{a(\tau)}f(\tau,x)f^\prime(\tau,x)\\
&+\frac{a^{\prime2}(\tau)}{2a^2(\tau)}f^2(\tau,x) - \frac{1}{2}\big(\nabla f(\tau,x)\big)^2 - a^2(\tau)V^{\prime\prime}(\phi)f^2(\tau,x)\bigg)\\
&=\int d\tau d^3x \frac{1}{2}\bigg(f^{\prime2}(\tau,x) -\big(\nabla f(\tau,x)\big)^2-2 \mathcal{H} f(\tau,x) f^\prime (\tau,x)\\
&+\mathcal{H}^2 f^2(\tau,x)-a^2(\tau) V^{\prime\prime}(\phi)f^2(\tau,x)\bigg)\\
\end{split}
\end{equation}

where $\mathcal{H}=\frac{a^\prime(\tau)}{a^2(\tau)}$ represents the Hubble parameter in terms of conformal time and $V^{\prime\prime} (\phi)$ is the second derivative of the potential $V(\phi)$ in terms of $\phi$. 

\begin{figure}
\centering
\includegraphics[width=1\textwidth]{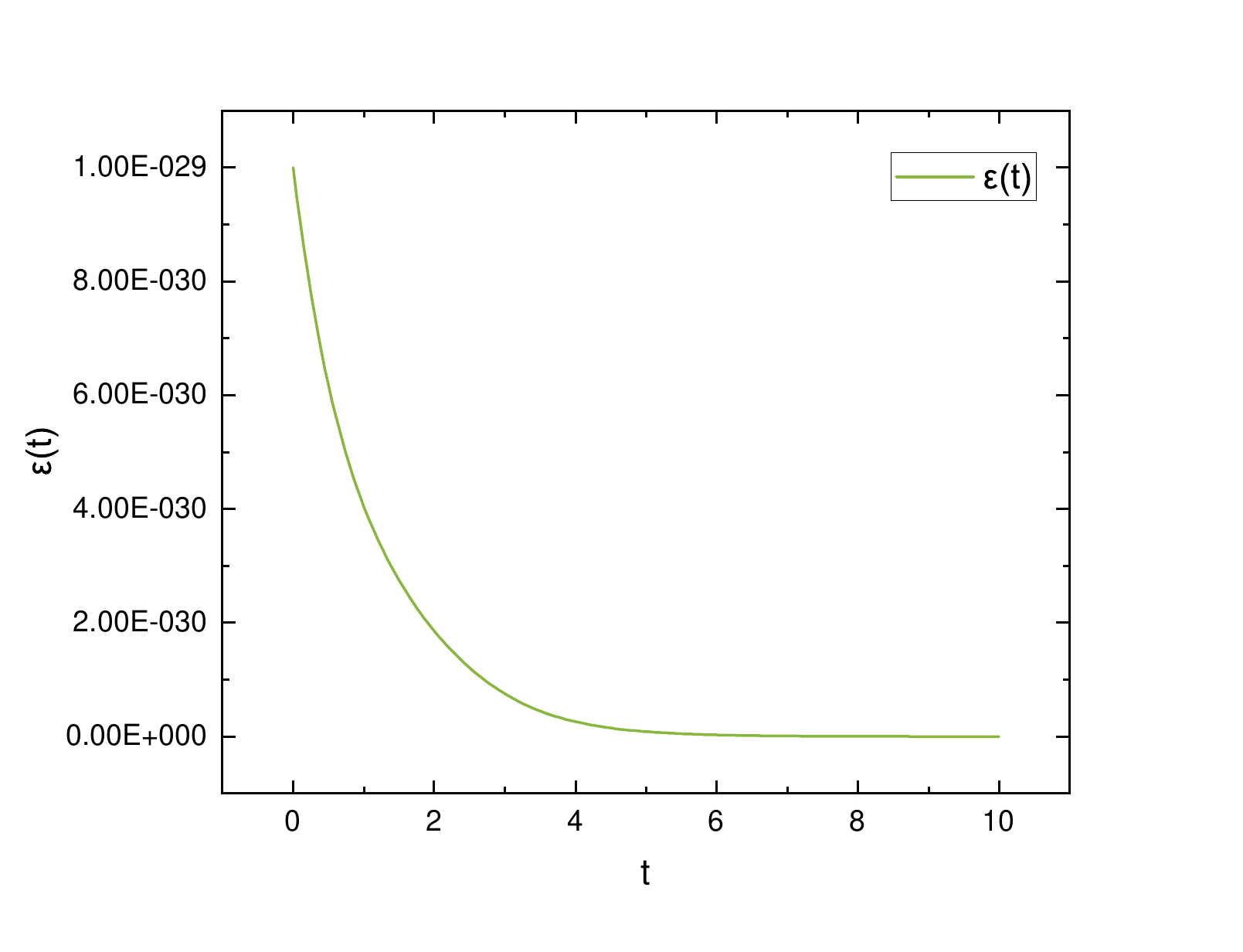}
\caption{Time evolution of the slow-roll parameter $\varepsilon$ defined in (\ref{GRepsilonNT}) in the presence of torsion. The value of  $\varepsilon$ at the end of inflation is within the range permitted by trans-Planckian censorship conjecture.}
\label{epsilonfig}
\end{figure}

\begin{figure}
\centering
\includegraphics[width=1\textwidth]{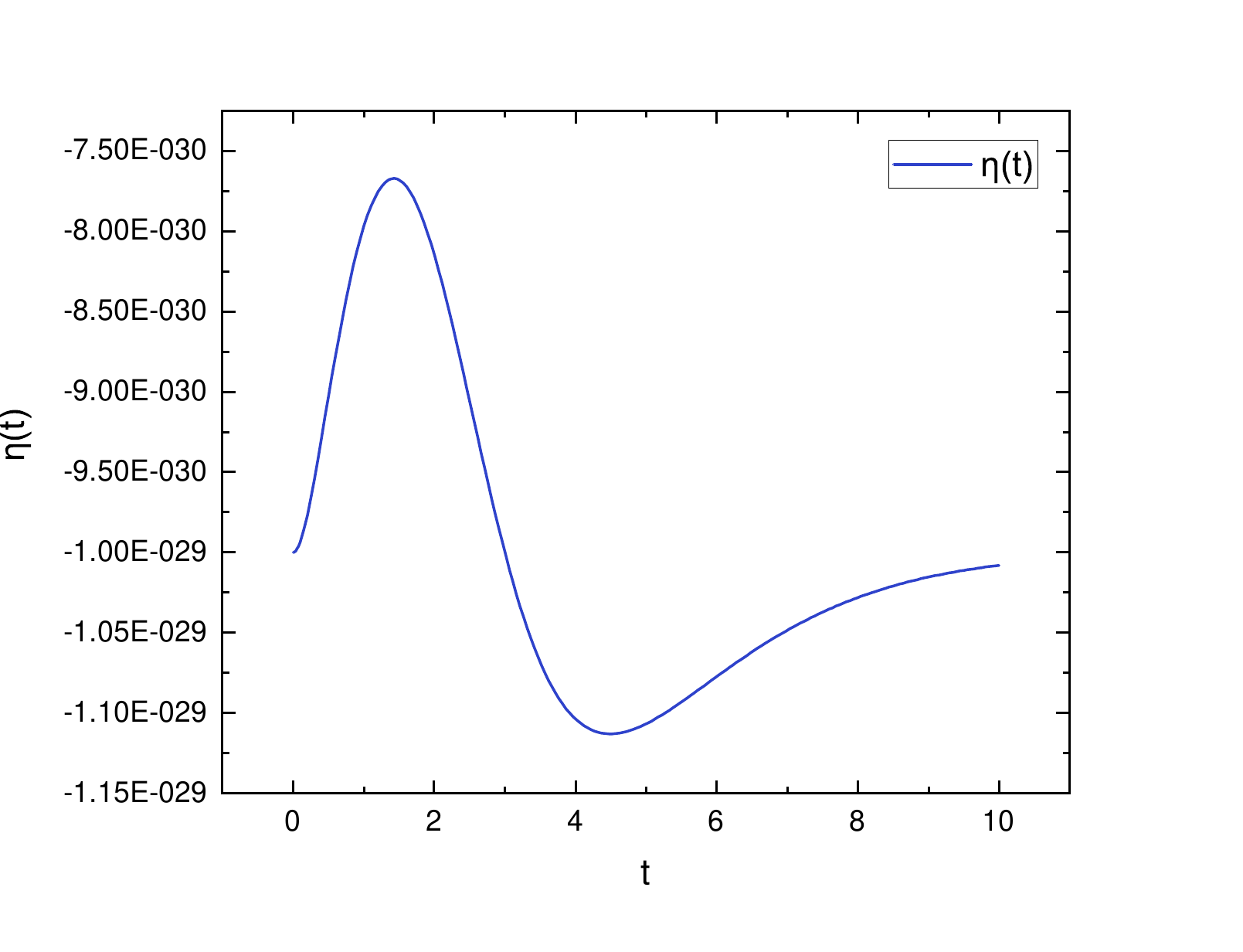}
\caption{Time evolution of the slow-roll parameter $\eta$ defined in (\ref{GRetaNT}) in the presence of torsion.}
\label{ettafig}
\end{figure}

The Hubble slow-roll parameters, which represent the condition for inflation, are defined as

\begin{equation}
\label{GRepsilonNT}
\varepsilon=-\frac{\dot{H}}{H^2}=1-\frac{\mathcal{H}^\prime}{\mathcal{H}^2}
\end{equation}

\begin{equation}
\label{GRetaNT}
\eta=\frac{\dot{\varepsilon}}{H \varepsilon}=\frac{a \varepsilon^\prime}{\mathcal{H} \varepsilon}
\end{equation}

Using the Friedmann equations (\ref{ECfriedmann}) and (\ref{ECAcc}) and also the relation between the spin parameter and the scale factor in the EC cosmology, i.e., equation (\ref{spinsquar}), one can derive the time evolution of these slow-roll parameters. Note that in effect in our model described by equations (\ref{ECfriedmann}) and (\ref{ECAcc}), there are two fields that contribute to the total energy density: The scalar field with $\omega \approx -1$ and fermionic matter with its own equation of state that acts as the source of spin and torsion in the universe. It should be noted that spin and torsion alone are capable of inducing inflation without the need for a scalar field, as demonstrated in \cite{Pereira, Gasp, Qorani}. In Einstein-Cartan theory, the spin induced inflation can only occur for a limited range of the equation of state parameter \cite{Gasp}, however such strict limitations are not present in the more general Poincar{\'e} gauge theory of gravity \cite{Qorani}. For a slow-rolling inflaton scalar field with $\omega \approx -1$ and a simple quadratic potential, the time evolution of $\varepsilon$ and $\eta$  in the presence of torsion are depicted in figures \ref{epsilonfig} and \ref{ettafig}. The figures show $\varepsilon,\vert\eta\vert \ll 1$ which indicates that we are in the slow-roll regime.  In \cite{BrandennVafa}, it has been shown that the TCC requires that, for quantum fluctuations of inflation to be responsible for the origin of the structure, there should be an upper bound to the value of the first slow-roll parameter $\varepsilon$ as $\varepsilon<10^{-31}$ at the end of inflation. As can be seen in the figure \ref{epsilonfig}, in our model in EC theory, the value of $\varepsilon$ can actually comply with the limitations of TCC at the end of inflation.

In order to examine the effects of torsion and spin on the scalar field perturbations, we assume that gravitational interactions at very high energies of the early universe where there is a high density of fermions can be described by Einstein-Cartan theory. As the spin density and its associated torsion were probably very large at the time of inflaton,  we will use the Friedmann equation in Einstein-Cartan theory (\ref{ECfriedmann}) to introduce the effects of spin and torsion into the quadratic action (\ref{quadAc}). By substituting the energy density of the inflaton field $\rho=\frac{1}{2} \dot{\phi}^2 +  V(\phi)$ into the Friedmann equation (\ref{ECfriedmann}) we get 

\begin{equation}
\label{M5}
\begin{split}
\mathcal{H}^2&= \frac{\kappa a^2(\tau)}{3} \bigg(\big(\frac{1}{2} \dot{\phi}^2 + V(\phi)\big)\,-\,\frac{1}{4}\kappa s^2 \bigg)\\
&=\frac{\kappa}{6} a^2(\tau) \dot{\phi}^2 + \frac{\kappa}{3}a^2(\tau) V(\phi) - \frac{\kappa}{12} \kappa a^2(\tau) s^2 \\
\end{split}
\end{equation}

And as a result, the inflaton potential $ V(\phi)$ can be expressed as 

\begin{equation}
 V(\phi)=\frac{3 \bigg(\mathcal{H}^2-\frac{\kappa}{6} a^2(\tau) \dot{\phi}^2 + \frac{\kappa^2}{12} s^2 a^2 (\tau) \bigg)} {a^2(\tau) \kappa}\\
\end{equation}

From the definition of potential slow-roll parameters we have the following

\begin{equation}
\label{M4}
\eta_V=M_{Pl}^2\big(\frac{V^{\prime\prime}(\phi)}{V(\phi)}\big)
\end{equation}

\begin{equation}
\label{VPhi}
 V^{\prime\prime}(\phi)=\frac{\eta_V V(\phi)}{M_{Pl}^2}
\end{equation}

Substituting (\ref{VPhi}) into (\ref{M5}), we finally get 

\begin{equation}
\label{Vpp}
V^{\prime\prime}(\phi)=\frac{V(\phi)\eta_V}{M^2_{Pl}}=\frac{\eta_\nu \bigg(\frac{3\mathcal{H}^2}{\kappa}-\frac{\dot{\phi}^2}{2}+\frac{\kappa s^2}{4 } \bigg)}{M^2_{Pl}}
\end{equation}

Equation (\ref{Vpp}) can be substituted into the last term in the quadratic action (\ref{quadAc}) and due to the presence of spin density parameter $s$, it carries the effects of torsion. We should note that the Hubble parameter is approximately constant during inflation and, also in the slow-roll regime, the kinetic term $\dot{\phi}^2$ is negligible. On the other hand, as stated earlier, the spin density $s$ was probably very high in the very early universe. As a result of these arguments, we can safely assume that during the inflationary period, the last term in (\ref{Vpp}) was dominant and we can safely neglect the other two terms. With these consideration the quadratic action will be

\begin{equation}
\label{M6}
\begin{split}
S^{(2)}\, & =\,\frac{1}{2} \int\,d\tau\,d^3x\,\bigg[f^{\prime 2} (\tau,x)\,-\,\big(\nabla f(\tau,x)\big)^2\\
&+\,\bigg(\frac{a^{\prime\prime}(\tau)}{a(\tau)}\, -\,a^2(\tau) V^{\prime\prime} (\phi)\bigg) f^2(\tau,x)\,\bigg]\\
& =\,\frac{1}{2} \int\,d\tau\,d^3x\,\bigg[f^{\prime 2} (\tau,x)\,-\,\big(\nabla f(\tau,x)\big)^2\\
&+\,\bigg(\frac{a^{\prime\prime}(\tau)}{a(\tau)}\, -\,a^2(\tau)\big(\frac{\eta_V \kappa s^2}{4M^2_{Pl}}\big)\bigg)\,f^2(\tau,x)\bigg]\\
\end{split}
\end{equation}

Now we can use the Euler-Lagrange equation 

\begin{equation}
\frac{\partial \mathcal{L}}{\partial f(\tau,x)}\,-\,\frac{d}{d\tau}\,\big(\frac{\partial \mathcal{L}}{\partial f'(\tau,x)}\big)\,=\,0,
\end{equation}

to derive the equation of motion for perturbation field. We have

\begin{equation}
\begin{split}
&\frac{\partial \mathcal{L}}{\partial f(\tau,x)}=2f(\tau,x) \bigg(\frac{a^{\prime \prime}(\tau)}{a(\tau)}-a^2(\tau) \big(\frac{\eta_V \kappa s^2}{4M^2_{Pl}} \big)\bigg)\\
&\frac{\partial \mathcal{L}}{\partial f^\prime(\tau,x)}=2f^{\prime}(\tau,x)
\end{split}
\end{equation}

Combining the above, we get the equivalent of the famous Mukhanov-Sasaki equation in Einstein-Cartan cosmology in the slow-roll regime

\begin{equation}
\label{M7}
f^{\prime\prime}(\tau,x)-\nabla^2f(\tau,x)-\bigg(\frac{a^{\prime\prime}(\tau)}{a(\tau)}-a^2(\tau) \big(\frac{\eta_V \kappa s^2 }{4M^2_{Pl}} \big)\bigg) f(\tau,x)=0
\end{equation}

where the effects of torsion are contained in the second term inside parentheses. In order to extract the time evolution of scalar perturbations, as usual, we perform a Fourier expansion on the perturbation parameter $f(\tau,x)$ as

\begin{equation}
\label{Fourier}
f(\tau,x)=\int \frac{d^3k}{(2\pi)^3} \exp^{ik.x} f_k(\tau)
\end{equation}

Substituting into (\ref{M7}) and simplifying, we have

\begin{equation}
\begin{split}
\label{F}
&f^{\prime\prime}_k(\tau)\ + \bigg(k^2-\frac{a^{\prime\prime}(\tau)}{a(\tau)}-a^2(\tau)\big(\frac{\eta_V \kappa s^2}{4 M^2_{Pl}}\big)\bigg) f_k(\tau)=0
\end{split}
\end{equation}

In order to solve the above equation, we relate the spin density parameter $s$ to the scale factor with the help of the relation (\ref{spinsquar}),  by doing so (\ref{F}) will become 

\begin{equation}
\label{P11}
f_k^{\prime \prime} (\tau)+\Big(k^2- \frac{a^{\prime \prime}}{a}-a^2 \big(\frac{\eta_V \kappa}{4M^2_{pl}}(B_\omega \rho_0^\frac{2}{1+\omega} a^{-6})\big) \Big)f_k(\tau)=0 \, .
\end{equation}

To solve the equation (\ref{P11}) and to obtain the time evolution of scalar field perturbation $f(\tau)$, we first examine the limiting regimes of this equation. At late times ($\tau \rightarrow 0$), the third term inside the parentheses of equation (\ref{P11}) is negligible compared to the other two terms, as it is proportional to $a^{-4}$. In this case, the equation is as follows

\begin{equation}
f_k(\tau)+\big(k^2- \frac{a^{\prime \prime}}{a}\big) f_k (\tau)=0
\end{equation}

if as usual we assume that the scale factor and conformal time have a relation as $a \sim \tau^{\frac{1}{2}-\mu}$, the solution to the above equation will be

\begin{equation}
f(\tau)=\sqrt{-\tau}\big(A\, J (\mu , -k\tau)+B\, Y(\mu,-k\tau)\big)
\end{equation}

where $J$ and $Y$ are Bessel functions of the first and second kind, respectively (the solution can also be written using Hankel functions) and $A$ and $B$ are constants which should be determined by initial conditions. On the other hand, in the very early universe ($\tau \rightarrow - \infty$)  the third term in (\ref{P11}) is dominant compared to the other two terms. In this case equation (\ref{P11}) becomes

\begin{equation}
\label{MS2}
f_k^{\prime \prime} (\tau)+\Big(k^2 - \big(\frac{\eta_V \kappa}{4M^2_{pl}}(B_\omega \rho_0^\frac{2}{1+\omega} a^{-4})\big) \Big)f_k(\tau)=0
\end{equation}

where the solution will be

\begin{equation}
\begin{split}
\label{P15}
f(\tau)&=C \sqrt{-\tau} \,J \Bigg(\frac{1}{4\mu},\frac{\sqrt{\frac{-\rho_{0}^{\frac{2}{1+\omega}} B_\omega \kappa \eta_V}{M_{pl}^2}}\,k\,\tau^{2\mu}}{4\mu} \Bigg) \\
&+ \, D \sqrt{-\tau} \,Y \Bigg(\frac{1}{4\mu},\frac{\sqrt{\frac{-\rho_{0}^{\frac{2}{1+\omega}} B_\omega \kappa \eta_V}{M_{pl}^2}}\,k\,\tau^{2\mu}}{4\mu} \Bigg)
\end{split}
\end{equation}

where again $C$ and $D$ must be determined by initial conditions. Note that the physical interval of the conformal time is in the form of $-\infty < \tau < 0$. The inside of the parentheses in (\ref{P15}) contains the coefficient of $B_\omega$ and therefore shows the effects of the spin tensor and thus the torsion on high energies.

\bigskip

\section{Torsion and Trans-Planckian Physics}\label{sec4}

Now we examine the issue of applying the initial conditions to the above solutions. Here we are interested in the effects of torsion at the very early universe where quantum effects were important. At early times during the inflation, all of the perturbation modes we are interested in were inside the horizon. On these scales, the fluctuations of the inflaton field are described by the use of quantum harmonic oscillators. The standard procedure in dealing with these oscillators involves expressing the perturbation field $f$ and its conjugate momentum $\pi=\frac{\partial \mathcal{L}}{\partial f'}$  where $ \mathcal{L}$ is the Lagrangian in the quadratic action (\ref{M6}) in terms of the ladder (or creation and annihilation) operators which obey standard commutation relations. The ladder operators here are time dependent, but they can be expressed in terms of operators at a fixed time by the help of Bogolubov transformations. The most important point in the procedure is the choice of the vacuum state, which is defined as the state annihilated by the lowering operator. This choice can be ambiguous in a time-dependent background. The standard choice is to impose the initial condition at $\tau \rightarrow - \infty$ which will lead to the unique Bunch-Davis vacuum \cite{Bunch}. However, as stated previously, due to our ignorance of the physics beyond the Planck scale, we cannot follow a mode that far back in conformal time. Instead to obtain the effects of torsion on trans-Planckian physics we use the method first introduced in \cite{Danielsson2} and impose the initial conditions in high energies at $\tau \rightarrow \frac{\Lambda}{Hk}$ instead of the usual $\tau \rightarrow -\infty$. Here $\Lambda$ is the energy scale (usually the Planck scale) in which the effects of trans-Plankian physics become important. Since the relationship between conformal time and the scale factor is in the form of $\tau=-\frac{1}{aH}$, the relationship between the comoving momentum $k$ and the physical momentum $p$ will be $k=ap=-\frac{p}{H}$.

\begin{figure}
    \centering
    \includegraphics[width=1\linewidth]{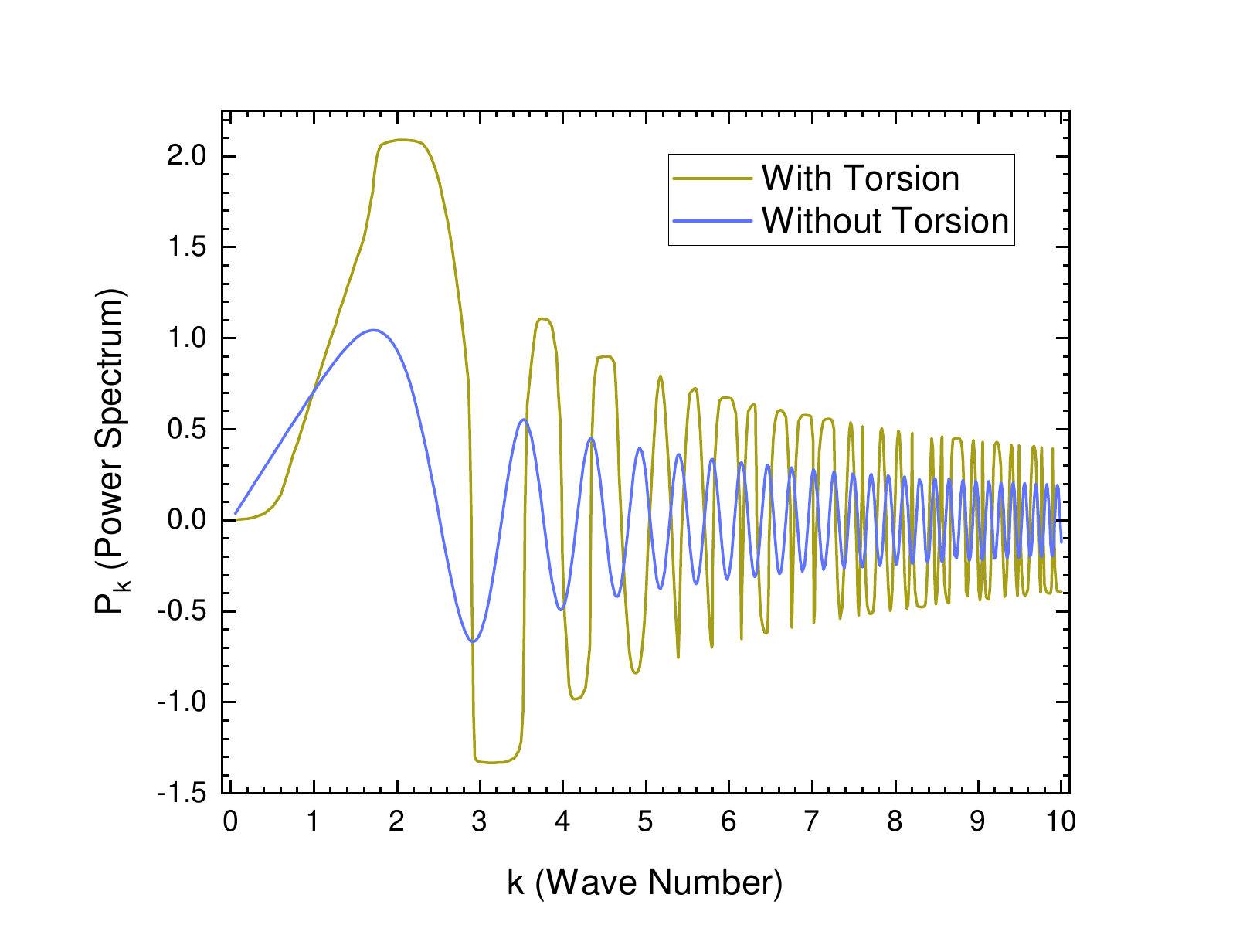}
    \caption{Power spectrum of scalar perturbations in terms of the wave number $k$ from numerically solving equation (\ref{P11}) in the absence of torsion (blue line) and in the presence of Torsion (green line). In both cases the new trans-Planckian initial conditions were imposed. In the cases with and without  torsion we set $B_{\omega}=1$ and $B_{\omega}=0$ in the numerical analysis of equation (\ref{P11}) respectively.}
    \label{Power Spectrum}
\end{figure}

Now we try to solve the equation (\ref{MS2}) with the initial condition imposed at  $\tau \rightarrow \frac{\Lambda}{Hk}$. In order to obtain the conformal time in which the initial conditions should be applied, we set the physical momentum $p$ equal to $\Lambda$. The solution to equation (\ref{P11}), with the new time for imposing initial conditions, is also a suitable combination of Bessel functions where the constant coefficients should be determined by the new initial conditions. Here, in order to show the trans-Planckian effects of torsion, we shall numerically solve equation (\ref{P11}) with the said initial conditions and determine the power spectrum for scalar perturbations in the presence of torsion. The scalar power spectrum is related to the perturbation function $f$ via the relation $P_k=\frac{k^3}{2 \pi^2} \vert f_k(\tau)\vert ^2$. In general relativity with the standard Bunch-Davis vacuum, imposing the initial conditions at $\tau \rightarrow -\infty$ leads to the standard relation $P_k=\Big(\frac{H}{2\pi}\Big)^2$, however, as found in \cite{Danielsson2}, imposing the initial condition at $\tau \rightarrow \frac{\Lambda}{Hk}$, causes the power spectrum $P_k$ to contain an oscillating term $P_k=\Big(\frac{H}{2\pi}\Big)^2 \Bigg(1-\frac{H}{\Lambda}\sin{\frac{2\Lambda}{H}}\Bigg)$. In the presence of torsion, the result for the trans-Planckian power spectrum is presented in figure \ref{Power Spectrum} after numerically solving equation (\ref{P11}) with the new initial conditions. The results show that the presence of torsion increases the amplitude of scalar perturbations for each given mode compared to the general relativistic result. As can be seen from the figure, when the initial conditions are imposed not at infinite past but at some other conformal time, the resulting power spectrum is no longer scale invariant. This and also the fluctuating behavior of the spectrum of scalar perturbations from the inflationary period due to the trans-Planckian effects is a phenomenon that has already been observed in previous work \cite{Danielsson2, Easther1}.

\bigskip
\section{Gravitational Waves from Inflation in The Presence of Torsion}\label{sec5}

We now turn our attention to the tensor perturbations generated after inflation. Here we follow the notation of reference \cite{Sasaki}.  To study the evolution of tensor perturbations, we consider a perturbed FRW metric in the form

\begin{equation}
\bar{g}_{\mu \nu}= g_{\mu \nu} + \delta g_{\mu \nu}
\end{equation}

and expand the tensor part of the metric perturbations in terms of the harmonic functions as

\begin{equation}
\begin{split}
\bar{g}_{00}&=a^{2}(\tau)\\
\bar{g}_{ij}& =-a^{2}(\tau)\Big(\gamma_{ij}+2 H_{\tau} Y_{ij}\Big) 
\end{split}
\end{equation}

Here, $\gamma_{ij}$ is the three-dimensional spatial metric for constant curvature hypersurfaces and $ Y_{ij}$ is the harmonic function. By perturbing the Einstein-Cartan field equation (\ref{EinsteinHehl}) and substituting this perturbed metric, we will have for the left-hand side

\begin{equation}
\Bar{\accentset{\circ}{G}}_{ij}=\accentset{\circ}{G}_{ij}+ \delta \accentset{\circ}{G}_{ij}
\end{equation}

where $ \accentset{\circ}{G}_{ij}$ is constructed fully by the Christoffel connection and so its perturbed components are well known (for example see appendix D in \cite{Sasaki}). For the right-hand side of equation (\ref{EinsteinHehl}), the relevant part of the perturbed standard energy-momentum tensor $\sigma_{\mu\nu}$ is given by

\begin{equation}
    \Bar{\sigma}_{ij} = p\Big[\delta_{ij}+\Pi_{T}Y_{ij}\Big]
\end{equation}

where $p$ is the isotropic pressure and $\Pi_{T}$ is the anisotropic stress. The rest of the terms on the right-hand side consist of the spin tensor (or equivalently torsion). Perturbing these term, we get

\begin{equation}
\delta \Tilde{\sigma}_{ij}=p\Pi_{T}Y_{ij}+ \kappa \Big[ 4 a^{-4} s \delta s - 6 a^{-2} s^2 H_{\tau} Y^{ij} \Big]
\end{equation}

where $\delta s$ is the perturbation in the spin density. As one can see from the above relations, the system involves perturbations of multiple fields and, in general, is quite complicated. In order to simplify the system of equations and estimate the effects of torsion on the spectrum of gravitational waves generated after inflation, we set $ \delta s= \Pi_{T}=0$. In this case, after doing a Fourier expansion of the perturbation field similar to the one in (\ref{Fourier}), the equation governing the time evolution of the tensor perturbation parameter $H_{\tau}$ can be derived from the perturbed field equation as

\begin{equation}
\label{12}
\begin{split}
H_\tau ^{\prime \prime} + 2 \frac{a^\prime}{a} H^\prime_\tau + \bigg[k^2 - 4 \frac{a^{\prime \prime}}{a} + & 2 (\frac{a^\prime}{a})^2 \bigg] H_\tau\\
 & = -6\,\kappa^2\, a^{-2}\,s^2 H_{\tau}  
 \end{split}
\end{equation}

\begin{figure}
    \centering
    \includegraphics[width=1\textwidth]{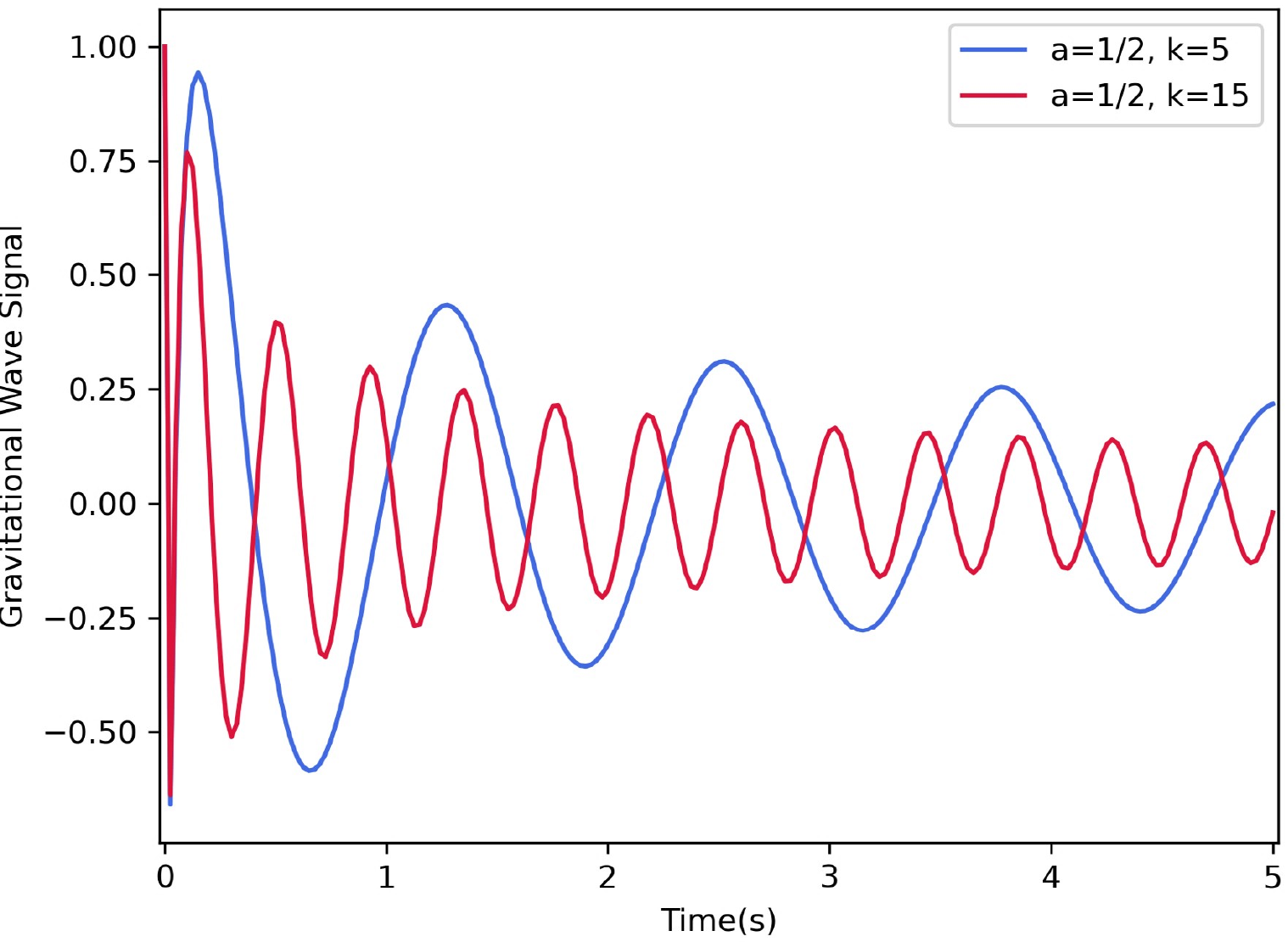}
    \caption{Gravitational wave signal from the numerical solution of equation (\ref{12}) for a radiation dominated universe with $a(t)\propto t^{\frac{1}{2}}$ and for different k numbers ($k=5,15$).}
    \label{GW1fig}
\end{figure}

\begin{figure}
    \centering
    \includegraphics[width=1\textwidth]{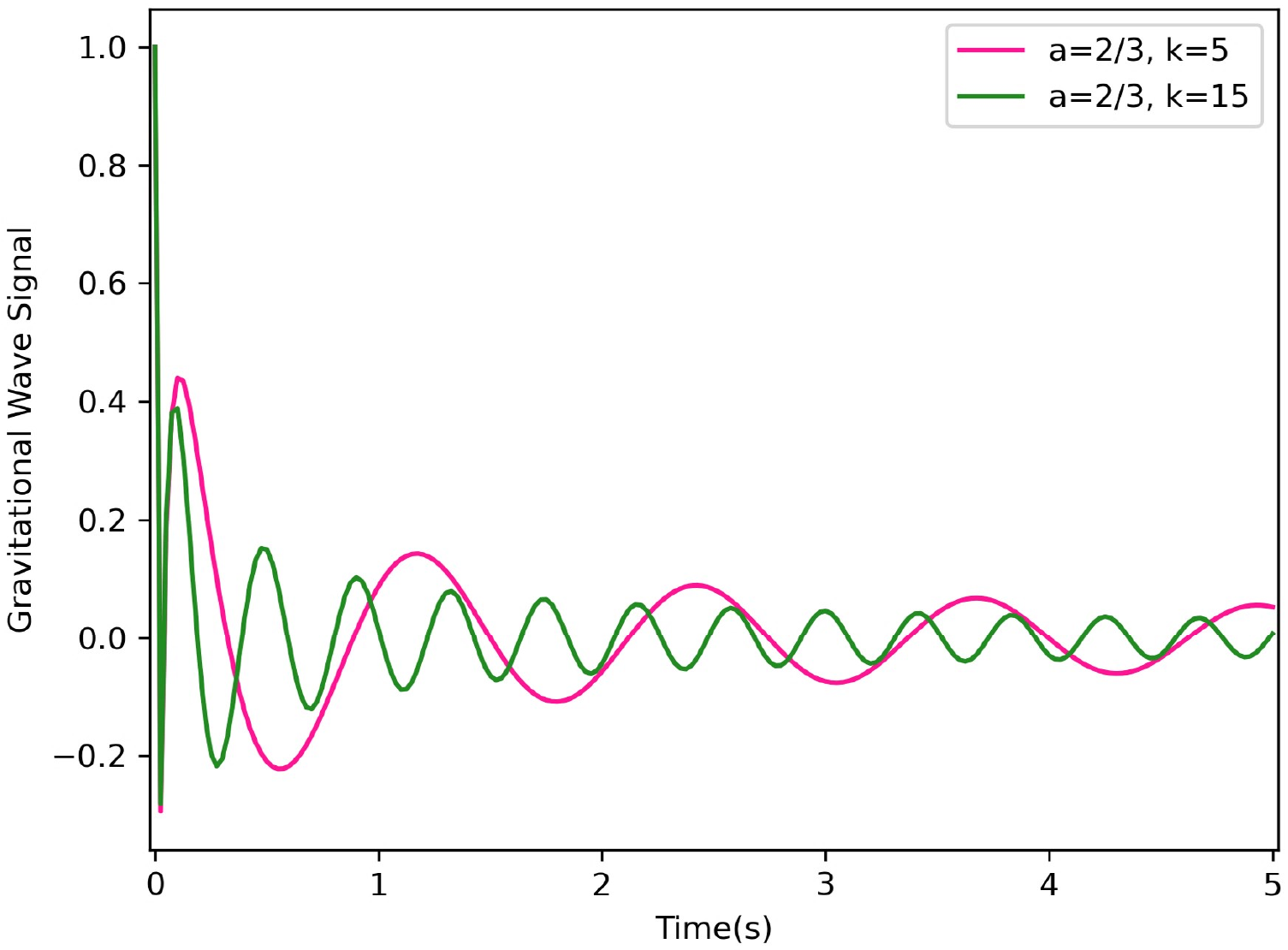}
    \caption{Gravitational wave signal from the numerical solution of equation (\ref{12}) for a matter dominated universe with $a(t)\propto t^{\frac{2}{3}}$ and for different wave numbers ($k=5,15$).}
    \label{GW2fig}
\end{figure}

\begin{figure}
    \centering
    \includegraphics[width=1\textwidth]{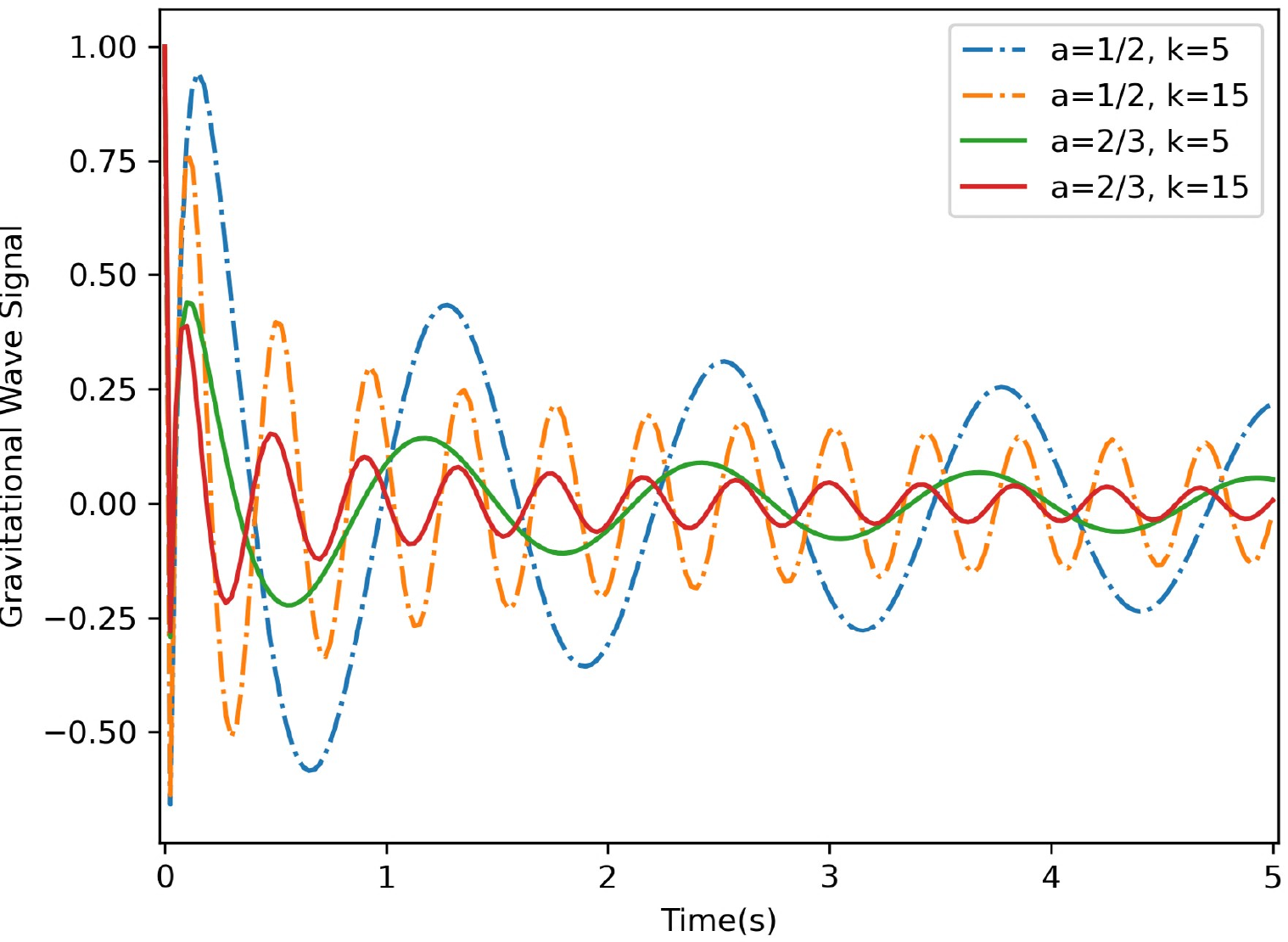}
    \caption{Gravitational wave signal for both radiation and matter dominated universes in the presence of torsion for different wave numbers ($k=5,15$).}
    \label{GW3fig}
\end{figure}

where again a prime denotes differentiation with respect to the conformal time. In the super-horizon scales, i.e., $\vert k\tau\vert \ll 1$, the term on the right-hand side of the equation (\ref{12}) can be neglected as it is proportional to $a^{-8}$. In this case, we can compute the spectrum of gravitational waves after the inflation by performing a first-order expansion in slow-roll parameters. By using equation (\ref{GRepsilonNT}), we can see that in this first-order expansion $\varepsilon$ should be a non-zero constant and

\begin{equation}
\label{14}
a(\tau)\propto \tau^{-(1+\varepsilon)}
\end{equation}

The proportionality constant can be fixed by normalizing $a(\tau)$ to the value that it has when a mode with a given wave number crosses the horizon. By doing this we get in the first order in slow-roll parameters \cite{Maggiore}
\begin{equation}
\label{15}
a(\tau)= (1+\varepsilon)\frac{k}{H}(-k\tau)^{-(1+\varepsilon)}
\end{equation}

Where here $H$ is the value of the Hubble parameter at the time of the horizon crossing. Substituting the above equation in (\ref{12}), the general solution in the super-horizon scales takes the form

\begin{equation}
\label{16}
\begin{split}
H_{\tau} & =E\, \tau^{\varepsilon+\frac{3}{2}} J \bigg(\frac{(12\varepsilon^2+44\varepsilon+33)^\frac{1}{2}}{2},k\tau \bigg) \\
&+ F\, \tau^{\varepsilon+\frac{3}{2}} Y \bigg(\frac{(12\varepsilon^2+44\varepsilon+33)^\frac{1}{2}}{2},k\tau \bigg)
\end{split}
\end{equation}

Where $E$ and $F$ again are constants to be determined by the initial conditions. The effects of torsion in this case are contained in the different values of Hubble and slow-roll parameters compared to the standard cosmology. Applying the initial conditions the same way as in the scalar case, i.e., at the finite conformal time $\tau \rightarrow \frac{\Lambda}{Hk}$, we get the same oscillating term for the spectrum of tensor perturbations. Using equation (\ref{15}), the spectrum is given by

\begin{equation}
P_T=\frac{k^3}{2\pi^2 a(\tau)^2} \vert H_{\tau}\vert^2
\end{equation}

where $a(\tau)$ is given by (\ref{15}). The tensor-to-scalar ratio $r$, defined as the ratio between amplitude of tensor and scalar perturbations then is given by

\begin{equation}
r=16\, \varepsilon+O(\varepsilon^2)
\end{equation}

From figure \ref{epsilonfig}, we can see that the value of the scalar-to-tensor ratio in our model is given by $r\approx 10^{-30}$, which is consistent with the value given in \cite{BrandennVafa} for any single or multiple field inflationary models complying with the TCC.

For the subsequent radiation and matter-dominated eras, we can write the spin density in equation (\ref{12}) in terms of the scale factor with the help of equation (\ref{spinsquar}) and then solve the equation numerically for different values of the equation of state parameter $\omega$. The result of the numerical analysis for two interesting cases of a radiation-dominated universe ($\omega=\frac{1}{3}$) and a matter-dominated universe ($\omega=0$) are given in figures \ref{GW1fig} and \ref{GW2fig}, respectively. Figure \ref{GW3fig} shows the gravitational wave signal for both matter- and radiation-dominated universes for different values of the wave number $k$.

\bigskip
\section{Conclusion}\label{sec6}
In this paper, considering torsion as a trans-Planckian effect, we studied the generation and propagation of both scalar and tensor perturbations during inflation. The key to accounting for the unknown trans-Planckian physics is to apply the initial conditions not at the infinite past but at a finite conformal time in which quantum effects become dominant. This amounts to a choice of vacuum different from the usual Bunch-Davis vacuum used in standard inflationary scenarios. The presence of torsion causes the amplitude of the trans-Planckian power spectrum to be greater than its general relativistic counterpart. Moreover, the trans-Planckian effects cause the power spectrum to no longer be scale-invariant.
As a further result, for a scalar inflaton field with a simple quadratic potential in the presence of torsion, the value of the first slow-roll parameter $\varepsilon$ lies within the range permitted by the TCC. We also found the tensor-to-scalar ratio in our model to be in the order $10^{-30}$, which implies that any detection of large primordial gravitational waves shows that they have a different origin than any inflationary model consistent with the TCC \cite{BedroyanVafa}. Here we considered Einstein-Cartan theory, the simplest of torsion theories, to study inflationary dynamics. A more comprehensive analysis can be done by using more general torsion theories of gravity with dynamical torsion, which can be considered in future works.
%%%%%%%%%%%%%%%%%%%%%%%%%%%%%%%%%%%%%%%%%
\section{Statements and Declarations}
\subsection{Funding}
No funding was received for conducting this study.
\subsection{Competing interests}
The authors have no relevant financial or non-financial interests to disclose.
\subsection{Data Availability}
Data sharing not applicable to this article as no datasets were generated or analysed during the current study.
\subsection{Author contribution statement}
Both authors contributed to the study conception and design. Both authors contributed equally to the calculations. The final manuscript text was written by Siamak Akhshabi. Elham Arabahmadi prepared all the figures. Both authors read and approved the final manuscript.
%%%%%%%%%%%%%%%%%%%%%%%%%%%%%%%%%%%%%%%%%

\end{document}